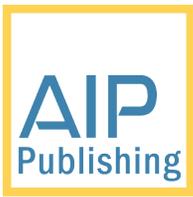

# Anisotropic Thermal Conductivity Measurement using A New Asymmetric-Beam Time-Domain Thermoreflectance (AB-TDTR) Method

**Man Li$^\parallel$, Joon Sang Kang$^\parallel$, Yongjie Hu\***

Department of Mechanical and Aerospace Engineering,

University of California, Los Angeles, Los Angeles, CA, 90095

$^\parallel$ M. Li and J.S. Kang contributed equally to this work.
\*Corresponding author. Email: yhu@seas.ucla.edu

**Abstract**

Anisotropic thermal properties are of both fundamental and practical interests, but remain challenging to characterize using conventional methods. In this work, a new metrology based on asymmetric beam time-domain thermoreflectance (AB-TDTR) is developed to measure three-dimensional anisotropic thermal transport by extending the conventional TDTR technique. Using an elliptical laser beam with controlled elliptical ratio and spot size, the experimental signals can be exploited to be dominantly sensitive to measure thermal conductivity along the cross-plane or any specific in-plane directions. An analytic solution for a multi-layer system is derived for the AB-TDTR signal in response to the periodical pulse, elliptical laser beam, and heating geometry, to extract the anisotropic thermal conductivity from experimental measurement. Examples with experimental data are given for various materials with in-plane thermal conductivity from 5 W/mK to 2000W/mK, including isotropic materials (silicon, boron phosphide, boron nitride), transversely isotropic materials (graphite, quartz, sapphire) and transversely anisotropic materials (black phosphorus). Furthermore, a detailed sensitivity analysis is conducted to guide the optimal setting of experimental configurations for different materials. The developed AB-TDTR metrology provides a new approach to accurately measure anisotropic thermal phenomena for rational materials design and thermal applications.

# I. Introduction

Anisotropic thermal transport is of both fundamental and practical importance. Orientation-dependent thermal conductivity has been observed in many materials systems because of their highly asymmetric crystal structures[1]. For example, the in-plane thermal conductivity of most familiar two-dimensional (2D) materials, e.g. graphene, hexagonal boron nitride, and molybdenum disulfide can be more than 10 times, even 100 times higher than their cross-plane thermal conductivity[2–8]. Furthermore, some 2D materials like black phosphorus have three-dimensional anisotropy, i.e., the in-plane thermal conductivity also depends on crystal orientation[8–14]. Importantly, the interaction between 2D lattices and external defects has been revealed to be highly anisotropic and phonon mode dependent through in situ thermal-electrochemical characterizations[8]. Polymers can also exhibit strong orientation dependent thermal conductivity, like more than 40 times larger thermal conductivity along chain direction than that of the transverse direction in polyethylene[15,16]. In addition to these van der Waals and covalent bonding mixed systems, materials with single form of bonding can also possess anisotropic thermal conductivity, like quartz, uranium dioxide, perovskites and so on[17–19]. Even materials considered as isotropic materials can display anisotropic thermal transport due to nonhomogeneous grains and defects during crystal growth, high aspect ratio nanostructure[20–22], superlattice or heterostructures[23], and measurement or device heating geometries[24–26]. For many applications, understanding the anisotropic thermal properties is a key merit to evaluate performance, e.g. thermal management in electronics[27,28], nuclear reactor design and safety[18,29], thermally stable photovoltaic conversion[19,30], directional thermoelectric conversion efficiency[31,32] and thermal regulation[8,33]. However, accurately measuring the anisotropic thermal transport remains challenging, despite that significant progress has been made recently.

The traditional spirit of measuring anisotropic thermal conductivity is to align temperature gradient and heat flux along the sample orientation that is of interest, so that no temperature gradient exists in any other directions, like steady-state methods and transient laser flash[34–37]. The prerequisite of big size samples or specified geometry limits their application for anisotropic measurement on novel materials. Modified 3ω-method and micro-bridge techniques can enable anisotropic thermal conductivity measurement on small samples[38–46], however, these methods usually require complicated microfabrications and materials processing to produce heating elements or resistive temperature sensors on the sample surface. Compared with micro-fabricated devices and 3ω method, the pump-probe optical spectroscopies such as time-domain thermoreflectance (TDTR), frequency-domain thermoreflectance, and transient thermal grating techniques can be used to perform non-contact and fast thermal conductivity measurement on both bulk and nanoscale samples[47–53], while isotropic heat conduction model was assumed in its early stage of development. To facilitate directional thermal conductivity measurement, TDTR has been recently modified to improve the measurement sensitivity to in-plane heat conduction[54–56]. Beam offset method was developed, but requires extra setups for translating beam positions and its data fitting under anisotropic heat conduction model can be time-consuming if the material is three-dimensional anisotropic[55]. Variable circular spot size has been applied but this method is not able to distinguish the in-plane anisotropy[56].

Here we develop a new metrology based on asymmetric beam time-domain thermoreflectance (AB-TDTR) to measure anisotropic thermal transport along cross-plane or any in-plane directions. The AB-TDTR signal sensitivity to thermal transport along different in-plane directions is decoupled and exploited using an elliptical laser beam with controlled elliptical ratio and spot size, to accomplish fast three-dimensional anisotropic thermal conductivity measurement. We first develop a mathematical model of anisotropic heat diffusion in multilayers heated by elliptical laser beam and provide the working principle of AB-TDTR method. Sensitivity analysis is conducted to guide the setting of experimental configuration for various materials. Demonstration experiments are performed on standard materials, including silicon (Si), cubic boron nitride (c-BN), boron phosphide, graphite, sapphire, quartz. Finally, the angle dependent thermal conductivity of black phosphorus is measured and compared with theoretical prediction to show the capability of our approach to measure thermal conductivity of three-dimensional anisotropic materials.

## II. Mathematical model and experimental principles

**A. Frequency domain solution to multilayer heat conduction equation with anisotropic thermal conductivity and asymmetric laser heating**

For pump probe spectroscopy measurement, a thin metal film is coated on the sample surface to serve as a thermal transducer, which is instantaneously heated by absorbing femtosecond pump laser pulses. This metal film also serves as the temperature sensor by reflecting a probe beam to photo diode with its reflectivity linearly proportional to temperature under a small temperature change. Thus, the transient temperature decay with time can be continuously detected by controlling the delay time between pump and probe beams using a mechanical stage and fitted with a multilayer thermal model to obtain the thermal conductivity (κ) of the sample. To achieve high signal to noise ratio, the laser pulses are modulated, and lock-in technique are used to detect temperature response at modulation frequency $f_0$. Under the assumption that temperature response of the sample to laser heating is both linear and time invariant, the detected signal by lock-in amplifier was given by Cahill as[52]

$$Z(f_0) = \frac{\beta Q Q_s}{T_s^2} \sum_{k=-\infty}^{\infty} H(2\pi f_0 + 2\pi f_s k) e^{2\pi i f_s k \tau} \quad (1)$$

where $H(f)$ is the frequency response of sample heated periodically, β indicating temperature coefficient of reflectivity and electronic gains, Q and $Q_s$ power of each pump pulse and probe pulse respectively, $T_s$ the period of laser pulses, $f_s$ the frequency of laser pulses, τ the delay time between pump beam and probe beam. The detected in-phase signal $V_{in}$ and out-of-phase signal $V_{out}$ by lock-in amplifier are real part and imaginary part of $Z(f_0)$. Since the phase signal $\tan^{-1}(V_{out}/V_{in})$ can exclude effects from some of the noise signals, we would discuss only the phase signals in the rest sections. For conventional TDTR, more details of experimental setup and derivation of $H(f)$ can be found in literatures[52,53]. As follows, the frequency response of anisotropic materials with elliptical laser heating is derived for analyzing AB-TDTR measurement signal.

Different from isotropic medium, where heat flux is always along temperature gradient direction, heat conduction in anisotropic medium is much more complicated, where heat flux is related with temperature gradient along all the directions. Mathematically, it can be expressed in cartesian coordinates as[17]

$$q_i = -\sum_{j=1}^{3} \kappa_{ij} \frac{\partial T}{\partial x_j} \quad (2)$$

where i, j mean the directions and T is temperature. Nine $\kappa_{ij}$ elements constitute a second rank tensor,

$$\kappa = \begin{bmatrix} \kappa_{xx} & \kappa_{xy} & \kappa_{xz} \\ \kappa_{yx} & \kappa_{yy} & \kappa_{yz} \\ \kappa_{zx} & \kappa_{zy} & \kappa_{zz} \end{bmatrix} \quad (3)$$

which is called thermal conductivity tensor. Under a certain orthogonal Cartesian coordinate system, the off-diagonal elements of thermal conductivity tensor can vanish[17]. The axes of this coordinate system are defined the principal thermal transport axes. The anisotropic heat conduction equation in AB-TDTR measurement in this coordinate is expressed as

$$\kappa_{xx}\frac{\partial^2 T}{\partial x^2} + \kappa_{yy}\frac{\partial^2 T}{\partial y^2} + \kappa_{zz}\frac{\partial^2 T}{\partial z^2} + S = C_v \frac{\partial T}{\partial t} \quad (4)$$

where S is the heat source term, $C_v$ volumetric heat capacity of anisotropic solids[17]. In the multi-layers model of AB-TDTR experiment, heat source term is zero. The laser heating term $I(x,y) = \frac{2A_0}{\pi w_{0,x} w_{0,y}} \exp(-\frac{2x^2}{w_{0,x}^2} - \frac{2y^2}{w_{0,y}^2})$ is treated as a heat flux boundary condition, where $A_0$ is the absorbed power of pump beam, $w_{0,x}$ and $w_{0,y}$ the $1/e^2$ semi-minor axis length and $1/e^2$ semi-major axis length respectively as shown in Fig. 1(a), which mean the distance from ellipse center to points on major axis and minor axis where laser intensity is $1/e^2$ of the peak intensity. In the following sections, the $1/e^2$ will not be explicitly given. To obtain frequency-domain solution of temperature response, heat conduction equation is written as after Fourier transformation

$$\frac{\partial^2 \hat{T}}{\partial z^2} = \lambda \hat{T} \quad (5)$$

$$\lambda = \frac{2\pi i f C_v}{\kappa_{zz}} + \frac{\kappa_{xx}}{\kappa_{zz}} \xi^2 + \frac{\kappa_{yy}}{\kappa_{zz}} \eta^2 \quad (6)$$

where $\hat{T}$ is temperature in frequency domain, $f$, $\xi$ and $\eta$ the variables in Fourier space corresponding to t, x and y. This one-dimensional multi-layer heat conduction equation was solved by Schmidt et al. following the approach of Carslaw and Jaegar[53,57]

$$\hat{T}(f,\xi,\eta) = \left(-\frac{D}{C}\right)\frac{A_0}{2\pi}\exp(-\frac{w_0^2\xi^2+w_s^2\eta^2}{8}) \quad (7)$$

where C and D are elements of transfer matrix as function of thickness and $\lambda$, indicating geometry and thermal properties of each layer[53]. After inverse Fourier transformation and taking the elliptical shape of probe beam into account, the frequency response function of AB-TDTR is derived as

$$H(f) = \frac{A_0 A_s}{4\pi^2}\iint_{-\infty}^{\infty}\left(-\frac{D}{C}\right)\exp(-\frac{(w_{0,x}^2+w_{s,x}^2)\xi^2+(w_{0,y}^2+w_{s,y}^2)\eta^2}{8})d\xi d\eta \quad (8)$$

where $A_s$ is the reflected power of probe beam, $w_{s,x}$ and $w_{s,y}$ the semi-minor length and semi-major length of probe beam. Here it should be noted that the only important characteristic sizes of laser beams in AB-TDTR are the root mean square of the minor axis length $D_x = 2\sqrt{\frac{w_{0,x}^2+w_{s,x}^2}{2}}$ and of the major axis length $D_y = 2\sqrt{\frac{w_{0,y}^2+w_{s,y}^2}{2}}$.

Note, the alignment between the elliptical beams and the principal crystal directions can be controlled using a rotating sample holder (Fig. 1a). Even under the coordinates that the asymmetric beams are not aligned with the principal crystal directions, when elliptical beams with very high elliptical ratio is used and cross-plane is along z axis, the heat conduction would be two-dimensional in x-z plane and the terms involving $\frac{\partial^2 T}{\partial x \partial y}$, $\frac{\partial^2 T}{\partial x \partial z}$ and $\frac{\partial^2 T}{\partial y \partial z}$ in heat conduction equation would also vanish[17]. The mathematical form of anisotropic heat conduction would stay unchanged as Eqn. (4). More discussion about anisotropic heat conduction will be given in the following for the demonstrative experiment on black phosphorus (BP).

**B. Experimental principles and sensitivity analysis**

The setup schematic of AB-TDTR is shown in Fig.1 (b). A pair of cylindrical lenses (THORLABS LJ1653L1-B and LK1419L1-B) and a pair of spherical lenses are combined to enable independently control of spot size ($D_x$ and $D_y$) along the x and y axis of the elliptical laser beams (Fig. 1c). The use of asymmetric beams instead of circular beams in AB-TDTR measurement enables the capability to precisely measure thermal conductivity along arbitrary directions of interest.

The key design of AB-TDTR is to decouple thermal transport along different directions. For conventional TDTR experiment, high modulation frequency and large beam spot size are usually used so that the thermal penetration depth ($L_p = \sqrt{\kappa/\omega_0 C_V}$) is much smaller than laser spot size. In this case, the temperature gradient is only along the cross-plane direction. By reducing the laser spot size to be close to or smaller than the in-plane $L_p$, in-plane thermal transport can affect the detection signal, however, it is not possible to distinguish thermal conductivity difference between different in-plane directions due to the circular beam symmetry. For AB-TDTR experiment, the $D_x$ and $D_y$ can be controlled to be close to and much longer than the in-plane $L_p$ respectively, so that the detected signal is dominantly sensitive to the heat transfer along $D_x$ (versus $D_y$) direction. Therefore, AB-TDTR enables the precise measurement of anisotropic in-plane thermal conductivity.

In addition, it should be noted that the elliptical modeling for AB-TDTR can minimize the measurement uncertainty from an imperfect laser beam profile. In conventional TDTR, circular beams are usually assumed in data analysis despite that the actual beam shape always deviates from a perfect circular shape. Such a deviation from a perfect circular beam to a practically elliptical beam can bring non-negligible errors for the fitted thermal conductivity. A hypothetical example is given below with the following listed parameters: isotropic thermal conductivity $\kappa$ = 100 W/mK, volumetric heat capacity $C_v$ = 2×10$^6$ J/m$^3$K, interfacial thermal conductance G = 1×10$^8$ W/m$^2$K. The actual elliptical ratio of is 1.25 and $D_x$ is 5 µm. Under modulation frequency of 1 MHz and a circular beam assumption, the fitted $\kappa$ would be 87 W/mK, 13 % lower than the actual value. Therefore, the anisotropic mathematical modeling presented for the AB-TDTR would avoid such an error to improve measurement accuracy.

To provide a guideline for optimizing experimental settings for different materials, a comprehensive sensitivity analysis is conducted in this work by varying materials anisotropy, beam spot size and modulation frequency. Considering the generality of this analysis, some properties are fixed at most common values. For example, the volumetric heat capacity is fixed at 2×10$^6$ J/m$^3$K since most of solid materials at room temperature have values from 1~3×10$^6$ J/m$^3$K[58], while G between our hypothetical sample and aluminum is 1×10$^8$ W/m$^2$K, located in the most common range of 1×10$^{7~9}$ W/m$^2$K[59]. The sensitivity of TDTR phase signal to a certain parameter α was defined as[60]

$$S_\alpha = \frac{\partial \ln(\theta)}{\partial \ln(\alpha)} \quad (9)$$

where α can be $\kappa_{xx}$, $\kappa_{yy}$. Here, a sensitivity ratio is defined as γ = $S_{\kappa_{xx}}/S_{\kappa_{yy}}$, representing the key metric to quantitatively determine the measurement uncertainty due to the two competing parameters, i.e. $\kappa_{xx}$, $\kappa_{yy}$. The sensitivity ratio as a functio of thermal conductivity and beam diameters are plotted as color contours in Fig. 2. These sensitivity plots clearly show that the $S\kappa_{xx}$ is much larger than $S\kappa_{yy}$ for most diameters and materials thermal conductivity, which verifies that the measurement signal is dominantly sensitive to $\kappa_{xx}$ so that $\kappa_{xx}$ can be precisely measuring by taking advantage of the elliptical beam settings. Furthermore, the sensitivity ratio shows a strong dependence on $D_x$, $D_y$, $\kappa_{xx}$, $\kappa_{yy}$ and modulation frequency $f_0$. As a general guidance, a big $D_x$ (with a fixed elliptical ratio), a high $f_0$, or a small $\kappa_{xx}$ is desirable in order to obtain the maximal measurement sensitivity to $\kappa_{xx}$. Although the effect of $\kappa_{zz}$ on $\gamma$ is relatively weak in comparison with other parameters, it shows that high cross-plane thermal conductivity increases sensitivity to $\kappa_{xx}$ due to an increased cross-plane thermal penetration depth and thereby an increased effective in-plane heating size in the deeper layer beneath the top surface. Overall, the sensitivity analysis verifies the sufficient measurement sensitivity of AB-TDTR method to measure the thermal conductivity of interest (i.e. $\kappa_{xx}$).

## III. Demonstrative AB-TDTR measurements on various materials

### A. Experimental details

Examples of AB-TDTR experiments are conducted on different materials from isotropic materials, 2D anisotropic materials, to 3D anisotropic materials. The optical setup of AB-TDTR is illustrated in Fig. 1 (b). In this setup, a Ti:Sapphire oscillator (Tsunami, Spectra-physics) generates a train of femtosecond laser pulses with 800 nm wavelength and 80 MHz repetition rate. A polarizing beam splitter divides the beam into a pump and a probe pulse. The pump beam is sinusoidally modulated by the electro-optic modulator (EOM) typically from 1 MHz to 20 MHz and fundamental frequency of pump beam was doubled by (BIBO), corresponding to the wavelength of 400 nm. Probe beam is delayed using mechanical delay stage from 0 to 6000 ps with a resolution of less than 1 ps. The signal is detected by a lock-in amplifier at modulation frequency after the photo diode converts the reflected probe beam intensity into electrical signal. Before the recombination of the pump and probe beams, four lenses are introduced to tune the size and shape of the pump beam, two of which are cylindrical lenses and the other two are spherical lenses as shown in Fig. 1 (b). By controlling the distance of the plano-concave and plano-convex lenses, the beam size can be manipulated. And the elliptical ratio can be tuned by controlling the distance between convex and concave cylindrical lens as shown in Fig. 1 (c). For all the samples in the following sections, 80nm aluminum films are coated on them to serve as transducer by using e-beam evaporator. To evaluate the measurement reliability, 10 measurements are performed at each sample condition for all the following experiments.

### B. Measurement of isotropic materials ($\kappa_{xx} = \kappa_{yy} = \kappa_{zz}$)

First, the AB-TDTR experiment is conducted on prototype isotropic materials, including silicon, cubic boron nitride, and cubic boron phosphide. As the first step, the cross-plane thermal conductivity $\kappa_{zz}$ is accurately measured. Based on the sensitivity analysis (Fig. 2), the experiment is first designed to achieve high sensitivity to the cross-plane thermal transport, for example, using a big beam spot size ($D_x$ ~30 μm) and high modulation frequency ($f_0$ = 9.8 MHz). As an example, the cross-plane thermal conductivity of the silicon sample was measured as $\kappa_{zz}$ = 138.2 ± 5.1 W/mK, consistent with literature[61,62]. Then by using smaller spot size $D_x$ = 7 μm and $D_y$ = 210 μm and small modulation frequency of $f_0$ = 1.1 MHz, the phase data with delay time from 500 ps to 5000 ps was probed as displayed in Fig. 3. By fitting with the thermal diffusion model described in section II. A, the in-plane thermal conductivity is measured consistent with $\kappa_{zz}$, within 10% uncertainty. This verifies that thermal conductivity in silicon is almost isotropic. We also applied our AB-TDTR to measure cubic boron nitride and boron phosphide and summarized the data in Fig. 6.

C. Measurement of transversely isotropic materials ($\kappa_{xx} = \kappa_{yy} \neq \kappa_{zz}$)

Next, AB-TDTR is performed on transversely isotropic material which possesses isotropic in-plane thermal conductivity but different from the cross-plane thermal conductivity. Highly oriented pyrolytic graphite (HOPG) is used as a prototype exemplary material here. Since in each layer of graphite carbon atoms are arranged at honeycomb lattice, thermal transport in basal plane is isotropic. A two-step measurement procedure is performed. First of all, the interfacial thermal conductance (G) and $\kappa_{zz}$ are measured at a high modulation frequency and a large beam spot. Specifically, here by using modulation frequency $f_0$ = 9.8 MHz and big circular spot with diameter of 30 μm, the interface conductance between graphite and aluminum is measured as 5.5×10$^7$ W/m$^2$K, which is consistent with the reported value at room temperature by Schmidt *et al*[64]. Second, AB-TDTR is applied with a low $f_0$ value for in-plane measurement. Specifically, the modulation frequency is set as $f_0$ = 1.1 MHz and beam spot size are $D_x$ = 30 μm and $D_y$ = 900 μm. Consequently, with the value of G fixed, $\kappa_{xx}$ was extracted by fitting the AB-TDTR measurement data (Fig. 4a). In addition, the angle dependent thermal conductivity of graphite is measured using AB-TDTR method by rotating sample around the laser incidence direction (Fig. 4b). $\kappa_{xx}$ and $\kappa_{zz}$ of graphite are measured as 2054.0±313.9 W/mK and 5.5±0.7 W/mK respectively with no angle dependence, consistent with literature data[53,54,56,65].

In addition to the highly anisotropic graphite, AB-TDTR is applied to measure weakly anisotropic materials with relatively low thermal conductivity such as quartz and sapphire used as examples. For quartz, because of the absolute value of thermal conductivity is less than 10 W/mK, smaller spot size $D_x$ = 3 μm is adopted to improve sensitivity to in-plane thermal conductivity. The AB-TDTR results on these relatively low thermal conductivity materials show consistency with literature (Fig.6), proving the applicability of our new metrology on relatively low thermal conductivity materials with small anisotropy[55,66]. In addition, we note that a modulation frequency ($f_0$) dependent $\kappa_{zz}$ was observed in transition metal chalcogenides[67]. To consider such an effect from modulation frequency dependence, $\kappa_{zz}$ is measured at the corresponding modulation frequency in order to extract $\kappa_{xx}$ from the AB-TDTR data.

**D. Measurement of transversely-anisotropic materials ($\kappa_{xx} \neq \kappa_{yy} \neq \kappa_{zz}$)**

As we mentioned in the previous section, the most important advantage of AB-TDTR over variable spot size approach[56] is the extended capability of measuring transversely anisotropic materials, in which the thermal conductivity show significant difference even in the transverse plane. Black phosphorus (BP) is an ideal material platform that shows strong three-dimensional isotropy due to its highly anisotropic lattice structure.

Here, we performed AB-TDTR measured and studied the angle dependent thermal conductivity of BP. The G between BP and aluminum is measured as $3.3 \times 10^7$ W/m²K by using a modulation frequency $f_0$ = 9.8 MHz and circular beam spot diameter of 30 μm. Measurement sensitivity analysis is simulated in Figure 5a and shows that the signal has sufficient measurement sensitivity to the thermal conductivity along zigzag or armchair direction when it is aligned with the major elliptical direction. AB-TDTR data of BP are displayed in Fig. 5b. Thermal conductivity are measured as 84.4±1.0 and 24.1±1.8 W/mK for zigzag and armchair direction respectively, in good agreement with the literature values[8–10,14].

Importantly, AB-TDTR measurement can clearly identify the diagonal elements in the thermal conductivity tensor, i.e., $\kappa_{xx}$, $\kappa_{yy}$, and $\kappa_{zz}$. Mathematically, the off-diagonal elements of thermal conductivity tensor as described by the anisotropic heat conduction equation[17,68]

$$\kappa_{xx}\frac{\partial^2 T}{\partial x^2} + \kappa_{yy}\frac{\partial^2 T}{\partial y^2} + \kappa_{zz}\frac{\partial^2 T}{\partial z^2} + 2\kappa_{xy}\frac{\partial^2 T}{\partial x \partial y} + 2\kappa_{xz}\frac{\partial^2 T}{\partial x \partial z} + 2\kappa_{yz}\frac{\partial^2 T}{\partial y \partial z} = C_v \frac{\partial T}{\partial t} \quad (10)$$

In AB-TDTR measurement, the temperature gradient (i.e., $2\kappa_{xy}\frac{\partial^2 T}{\partial x \partial y}$) along major direction is vanishing, under a large elliptical ratio of laser beams. When the incidence direction of laser beam is normal to one principal directions, the other two off-diagonal terms would also vanish, and the heat conduction equation would become

$$\kappa_{xx}\frac{\partial^2 T}{\partial x^2} + \kappa_{yy}\frac{\partial^2 T}{\partial y^2} + \kappa_{zz}\frac{\partial^2 T}{\partial z^2} = C_v \frac{\partial T}{\partial t} \quad (11).$$

Based on equation (11), we can calculate the new diagonal elements of thermal conductivity when we rotate an angle θ between zigzag direction of BP and the minor axis direction of beam. First, we can build a new coordinate by rotating θ degree along z axis. In the new coordinate, x'cos θ = x and x'sin θ = y, where x' is along the minor axis direction of laser beams, x along zigzag direction and y along armchair direction. The new heat conduction equation would be

$$\kappa_{xx}cos^2\theta \frac{\partial^2 T}{\partial x'^2} + \kappa_{yy}sin^2\theta \frac{\partial^2 T}{\partial x'^2} + \kappa_{zz}\frac{\partial^2 T}{\partial z^2} = C_v \frac{\partial T}{\partial t} \quad (12)$$

So the thermal conductivity element along $x'$ would be

$$k_{x'x'} = k_{xx}cos^2\theta + k_{yy}sin^2\theta \quad (13)$$

where $\kappa_{xx}$ and $\kappa_{yy}$ are thermal conductivity along zigzag and armchair direction respectively. Based on Eqn. (13), the angle-dependent diagonal elements of thermal conductivity tensor are predicted (Fig. 5c). Experimentally measured thermal conductivity using AB-TDTR as a function of angle is plotted together and shows great agreement with the model prediction (Fig. 5c). This suggests that our AB-TDTR measurement result is a direct representation of the diagonal elements.

## Conclusion

In summary, a new metrology based on AB-TDTR method is developed for anisotropic thermal measurement. Experiments are conducted for different materials with a wide range of thermal conductivity values from ~ 5 to 2000 W/mK and thermal conductivity anisotropy from 0.5 to 400. The AB-TDTR measurement results are plotted in Fig. 6 and show good agreement with the literature. This study proves AB-TDTR method as a new metrology to precisely measure anisotropic thermal properties. This development enables a powerful platform to characterize advanced thermal materials and better understand thermal transport mechanisms.

## Acknowledgements

Y.H. acknowledges support from a CAREER award from the National Science Foundation under grant 1753393, a Young Investigator Award from the United States Air Force Office of Scientific Research under grant FA9550-17-1-0149, a PRF Doctoral New Investigator Award from the American Chemical Society under grant 58206-DNI5, the UCLA Sustainable LA Grand Challenge, and the Anthony and Jeanne Pritzker Family Foundation.

# References

[1] D.G. Cahill, P. V. Braun, G. Chen, D.R. Clarke, S. Fan, K.E. Goodson, P. Keblinski, W.P. King, G.D. Mahan, A. Majumdar, H.J. Maris, S.R. Phillpot, E. Pop, and L. Shi, Appl. Phys. Rev. **1**(1), 11305 (2014).

[2] Y.S. Touloukian, R.W. Powel, C.Y. Ho, and P.G. Klemens, in *Thermophys. Prop. Matter* (IFI/Plenum, New York, 1970), p. 1389.

[3] W. Jang, Z. Chen, W. Bao, C.N. Lau, and C. Dames, Nano Lett. **10**(10), 3909 (2010).

[4] S. Ghosh, W. Bao, D.L. Nika, S. Subrina, E.P. Pokatilov, C.N. Lau, and A.A. Balandin, Nat. Mater. **9**(7), 555 (2010).

[5] J.H. Seol, I. Jo, A.L. Moore, L. Lindsay, Z.H. Aitken, M.T. Pettes, X. Li, Z. Yao, R. Huang, D. Broido, N. Mingo, R.S. Ruoff, and L. Shi, Science **328**(5975), 213 (2010).

[6] L. Lindsay, W. Li, J. Carrete, N. Mingo, D.A. Broido, and T.L. Reinecke, Phys. Rev. B **89**(15), 155426 (2014).

[7] C. Chiritescu, D.G. Cahill, N. Nguyen, D. Johnson, A. Bodapati, P. Keblinski, and P. Zschack, Science **315**(5810), 351 (2007).

[8] J.S. Kang, M. Ke, and Y. Hu, Nano Lett. **17**(3), 1431 (2017).

[9] H. Jang, J.D. Wood, C.R. Ryder, M.C. Hersam, and D.G. Cahill, Adv. Mater. **27**(48), 8017 (2015).

[10] B. Sun, X. Gu, Q. Zeng, X. Huang, Y. Yan, Z. Liu, R. Yang, and Y.K. Koh, Adv. Mater. **29**(3), 1603297 (2017).

[11] B. Smith, B. Vermeersch, J. Carrete, E. Ou, J. Kim, N. Mingo, D. Akinwande, and L. Shi, Adv. Mater. **29**(5), 1603752 (2017).

[12] S. Lee, F. Yang, J. Suh, S. Yang, Y. Lee, G. Li, H.S. Choe, A. Suslu, Y. Chen, C. Ko, J. Park, K. Liu, J. Li, K. Hippalgaonkar, J.J. Urban, S. Tongay, and J. Wu, Nat. Commun. **6**, 8573 (2015).

[13] Z. Luo, J. Maassen, Y. Deng, Y. Du, R.P. Garrelts, M.S. Lundstrom, P.D. Ye, and X. Xu, Nat. Commun. **6**, 8572 (2015).

[14] J. Zhu, H. Park, J.Y. Chen, X. Gu, H. Zhang, S. Karthikeyan, N. Wendel, S.A. Campbell, M. Dawber, X. Du, M. Li, J.P. Wang, R. Yang, and X. Wang, Adv. Electron. Mater. **2**(5), 1600040 (2016).

[15] S. Shen, A. Henry, J. Tong, R. Zheng, and G. Chen, Nat. Nanotechnol. **5**(4), 251 (2010).

[16] X. Wang, V. Ho, R.A. Segalman, and D.G. Cahill, Macromolecules **46**(12), 4937 (2013).

[17] D.W. Hahn and M.N. Özişik, *Heat Conduction in Anisotropic Solids* (Wiley Online Library, 2012), p.614-650.

[18] K. Gofryk, S. Du, C.R. Stanek, J.C. Lashley, X.Y. Liu, R.K. Schulze, J.L. Smith, D.J. Safarik, D.D. Byler, K.J. McClellan, B.P. Uberuaga, B.L. Scott, and D.A. Andersson, Nat. Commun. **5**, 4551 (2014).

[19] M. Wang and S. Lin, Adv. Funct. Mater. **26**(29), 5297 (2016).

[20] J.E. Graebner, S. Jin, G.W. Kammlott, J.A. Herb, and C.F. Gardinier, Nature **359**(6394), 401 (1992).

[21] A. Sood, J. Cho, K.D. Hobart, T.I. Feygelson, B.B. Pate, M. Asheghi, D.G. Cahill, and K.E. Goodson, J. Appl. Phys. **119**(17), 5103 (2016).

[22] Z. Aksamija and I. Knezevic, Phys. Rev. B - Condens. Matter Mater. Phys. **82**(4),

5319 (2010).

[23] M.N. Luckyanova, J.A. Johnson, A.A. Maznev, J. Garg, A. Jandl, M.T. Bulsara, E.A. Fitzgerald, K.A. Nelson, and G. Chen, Nano Lett. **13**(9), 3973 (2013).

[24] J.S. Kang, H. Wu, and Y. Hu, Nano Lett. **17**(12), 7507 (2017).

[25] Y. Hu, L. Zeng, A.J. Minnich, M.S. Dresselhaus, and G. Chen, Nat. Nanotechnol. **10**(8), 701 (2015).

[26] Z. Chen and C. Dames, Appl. Phys. Lett. **107**(19), 193104 (2015).

[27] Y. Won, J. Cho, D. Agonafer, M. Asheghi, and K.E. Goodson, IEEE Trans. Components, Packag. Manuf. Technol. **5**(6), 737 (2015).

[28] A. Bar-Cohen and P. Wang, in *Nano-Bio- Electron. Photonic MEMS Packag.* (2010), pp. 349–429.

[29] N. Ghoniem and D. Walgraef, *Instabilities and Self-Organization in Materials* (Oxford Univ. Press, 2008),.

[30] D. Kraemer, B. Poudel, H.P. Feng, J.C. Caylor, B. Yu, X. Yan, Y. Ma, X. Wang, D. Wang, A. Muto, K. McEnaney, M. Chiesa, Z. Ren, and G. Chen, Nat. Mater. **10**(7), 532 (2011).

[31] C. Dames, S. Chen, C.T. Harris, J.Y. Huang, Z.F. Ren, M.S. Dresselhaus, and G. Chen, Rev. Sci. Instrum. **78**(10), 104903 (2007).

[32] X. Yan, B. Poudel, Y. Ma, W.S. Liu, G. Joshi, H. Wang, Y. Lan, D. Wang, G. Chen, and Z.F. Ren, Nano Lett. **10**(9), 3373 (2010).

[33] G. Wehmeyer, T. Yabuki, C. Monachon, J. Wu, and C. Dames, Appl. Phys. Rev. **4**(4), 41304 (2017).

[34] G.A. Slack, Phys. Rev. **122**(5), 1451 (1961).

[35] G.A. Slack, Phys. Rev. **127**(3), 694 (1962).

[36] W.J. Parker, R.J. Jenkins, C.P. Butler, and G.L. Abbott, J. Appl. Phys. **32**(9), 1679 (1961).

[37] M. Golombok and L.C. Shirvill, J. Appl. Phys. **63**(6), 1971 (1988).

[38] D.G. Cahill, Rev. Sci. Instrum. **61**(2), 802 (1990).

[39] Y.S. Ju, Thin Solid Films **339**(1999).

[40] C. Dames and G. Chen, Rev. Sci. Instrum. **76**(12), 1 (2005).

[41] P. Kim, L. Shi, A. Majumdar, and P.L. McEuen, Phys. Rev. Lett. **87**(21), 215502 (2001).

[42] D. Li, Y. Wu, P. Kim, L. Shi, P. Yang, and A. Majumdar, Appl. Phys. Lett. **83**(14), 2934 (2003).

[43] V. Mishra, C.L. Hardin, J.E. Garay, and C. Dames, Rev. Sci. Instrum. **86**(5)(2015).

[44] Y.M. Brovman, J.P. Small, Y. Hu, Y. Fang, C.M. Lieber, and P. Kim, J. Appl. Phys. **119**(23)(2016).

[45] J. Zheng, M.C. Wingert, E. Dechaumphai, and R. Chen, Rev. Sci. Instrum. **84**(11), 114901 (2013).

[46] Y. Zeng and A. Marconnet, Rev. Sci. Instrum. **88**(4)(2017).

[47] D. A. Young, C. Thomsen, H.T. Grahn, H.J. Maris, and J. Tauc, *Phonon Scattering in Condensed Matter*, edited by A. C. Anderson and J. P. Wolfe (Springer, Berlin, 1986), p.49.

[48] C.A. Paddock and G.L. Eesley, J. Appl. Phys. **60**(1), 285 (1986).

[49] W.S. Capinski and H.J. Maris, Rev. Sci. Instrum. **67**(8), 2720 (1996).

[50] B. Bonello, B. Perrin, and C. Rossignol, J. Appl. Phys. **83**(6), 3081 (1998).

[51] N. Taketoshi, T. Baba, E. Schaub, and A. Ono, Rev. Sci. Instrum. **74**(12), 5226 (2003).

[52] D.G. Cahill, Rev. Sci. Instrum. **75**(12), 5119 (2004).

[53] A.J. Schmidt, X. Chen, and G. Chen, Rev. Sci. Instrum. **79**(11), 114902 (2008).

[54] J.P. Feser and D.G. Cahill, Rev. Sci. Instrum. **83**(10), 104901 (2012).

[55] J.P. Feser, J. Liu, and D.G. Cahill, Rev. Sci. Instrum. **85**(10), 104903 (2014).

[56] P. Jiang, X. Qian, and R. Yang, Rev. Sci. Instrum. **88**(7), 74901 (2017).

[57] H.S. Carslaw and J.C. Jaeger, *Conduction of Heat in Solids* (Oxford University Press, New York, 1959), p.109.

[58] T.L. Bergman and F.P. Incropera, *Fundamentals of Heat and Mass Transfer* (John Wiley & Sons, 2011),.

[59] D.G. Cahill, W.K. Ford, K.E. Goodson, G.D. Mahan, A. Majumdar, H.J. Maris, R. Merlin, and S.R. Phillpot, J. Appl. Phys. **93**(2), 793 (2003).

[60] B.C. Gundrum, D.G. Cahill, and R.S. Averback, Phys. Rev. B - Condens. Matter Mater. Phys. **72**(24)(2005).

[61] H.R. Shanks, P.D. Maycock, P.H. Sidles, and G.C. Danielson, Phys. Rev. **130**(5), 1743 (1963).

[62] R.B. Wilson and D.G. Cahill, Nat. Commun. **5**, 5075 (2014).

[63] Kang *et al*., manuscript submitted.

[64] A.J. Schmidt, K.C. Collins, A.J. Minnich, and G. Chen, J. Appl. Phys. **107**(10), 104907 (2010).

[65] E. David R Lide, Taylor Fr. Boca Rat. FL , 0 (2007).

[66] E.R. Dobrovinskaya, L.A. Lytvynov, and V. Pishchik, *Sapphire: Material,*

*Manufacturing, Applications* (2009), p.110.

[67] P. Jiang, X. Qian, X. Gu, and R. Yang, Adv. Mater. **29**(36)(2017).

[68] J.M. Powers, J. Heat Transfer **126**(5), 670 (2004).

[69] J. A. Malen, K. Baheti, T. Tong, Y. Zhao, J. A. Hudgings, and A. Majumdar, J. Heat Transfer. **133**(8), 081601 (2011).

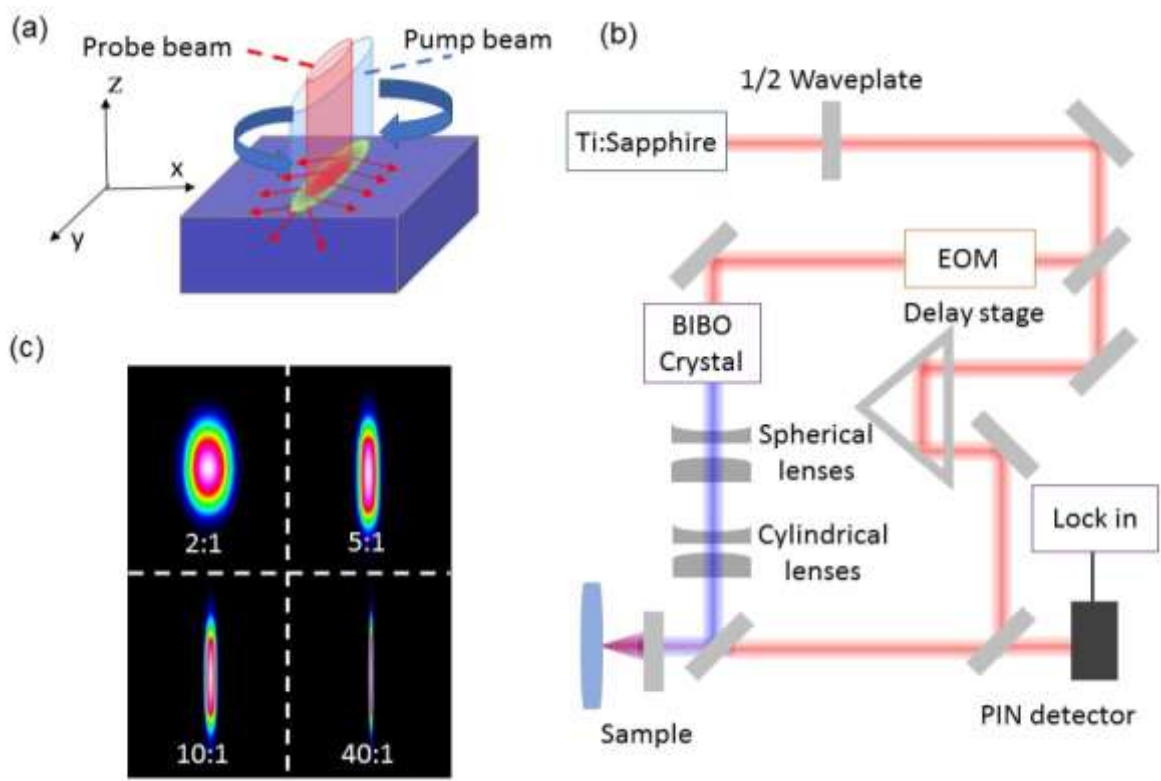

Figure 1. Working principles of the asymmetric-beam time-domain thermoreflectance technique (AB-TDTR) for measuring anisotropic thermal conductivity. (a) Schematic illustrates the alignment of elliptical beams to the sample surface using a rotation sample holder and the resulted temperature distribution due to laser heating. The major principal axis (y) and minor axis (x) represents the direction with the longest and shortest diameter of the laser beam. To measure anisotropic thermal conductivity, the heat conduction direction of interest is aligned to be in parallel with the y axis. (b) Schematic of the AB-TDTR setup. A pair of cylindrical lenses and a pair of spherical lenses are used to control the elliptical axis lengths ($D_x$ and $D_y$) of pump and probe beams independently. (c) Beam profiler images of example laser beams with different elliptical ratios.

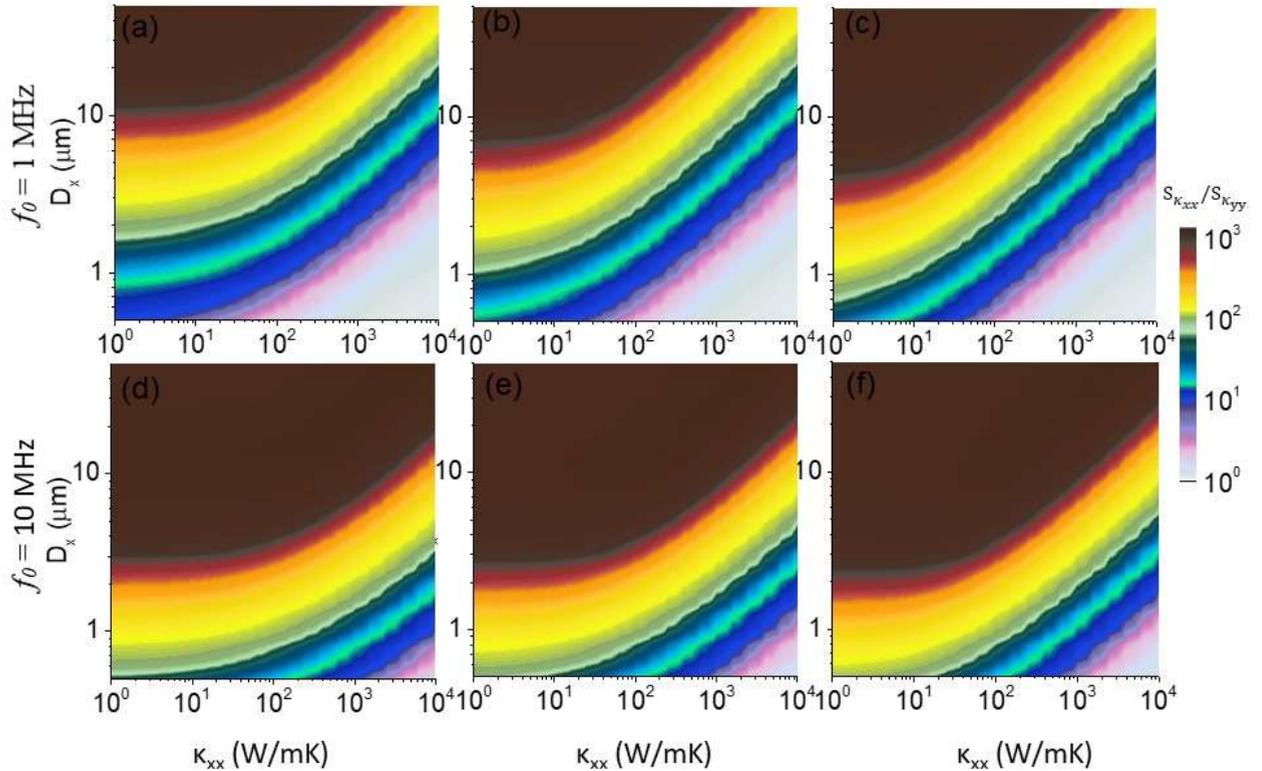

Figure 2. Sensitivity contours for AB-TDTR setting design of the modulation frequency, the laser spot size ($D_x$, $D_y$) considering in-plane isotropic materials with different thermal conductivity and anisotropy. Sensitivity analysis are conducted for different materials with $\kappa_{zz}$= 1, 10, and 100 W/mK (left to right) and with the modulation frequency of 1 and 10 MHz (top to bottom) under a fixed beam elliptical ratio of 40. The amplitude of the color scale is defined as the sensitivity ratio of phase signal, i.e. $S_{\kappa_{xx}}/S_{\kappa_{yy}}$, and calculated at a fixed delay time of 100 ps. The sensitivity contours verify the high measurement sensitivity to the thermal conductivity along the direction of interest, i.e. $\kappa_{xx}$. The strong contrast in sensitivity ratio provide guidance for optimizing the measurement accuracy.

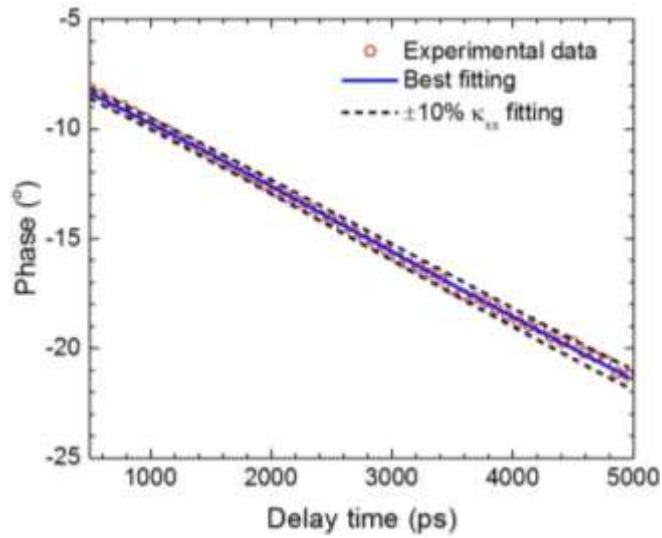
Figure 3. A typical AB-TDTR measurement data set (circles) for silicon, along with the best fitting curve (blue line) from the thermal diffusion model. Calculated curves (dash lines) with ±10% variation of the best fitting value $\kappa_{zz}$ are plotted to show the sensitivity. The modulation frequency is set as 1.1 MHz and beam size is $D_x = 7$ μm, $D_y = 210$ μm.

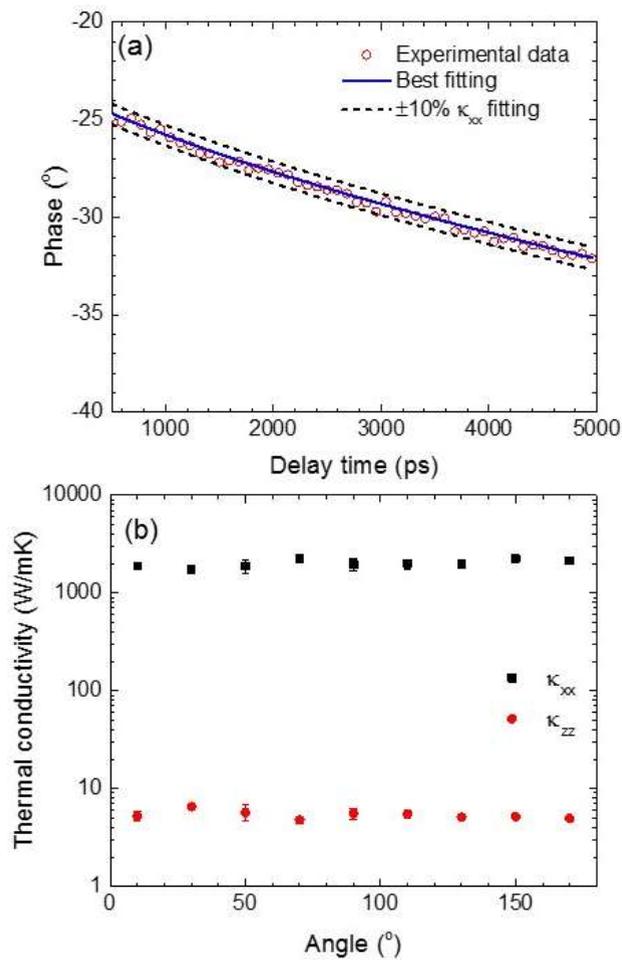

Figure 4. AB-TDTR measurement of angle-dependent thermal conductivity of graphite. (a) A typical AB-TDTR phase data set for graphite, along with the best fitting curve and ±10% $\kappa_{zz}$ fitting curves. The modulation frequency is set as 1.1 MHz and beam size is $D_x$ = 30 μm, $D_y$ = 900 μm. (b) Angle dependence of in-plane and cross-plane thermal conductivity of graphite.

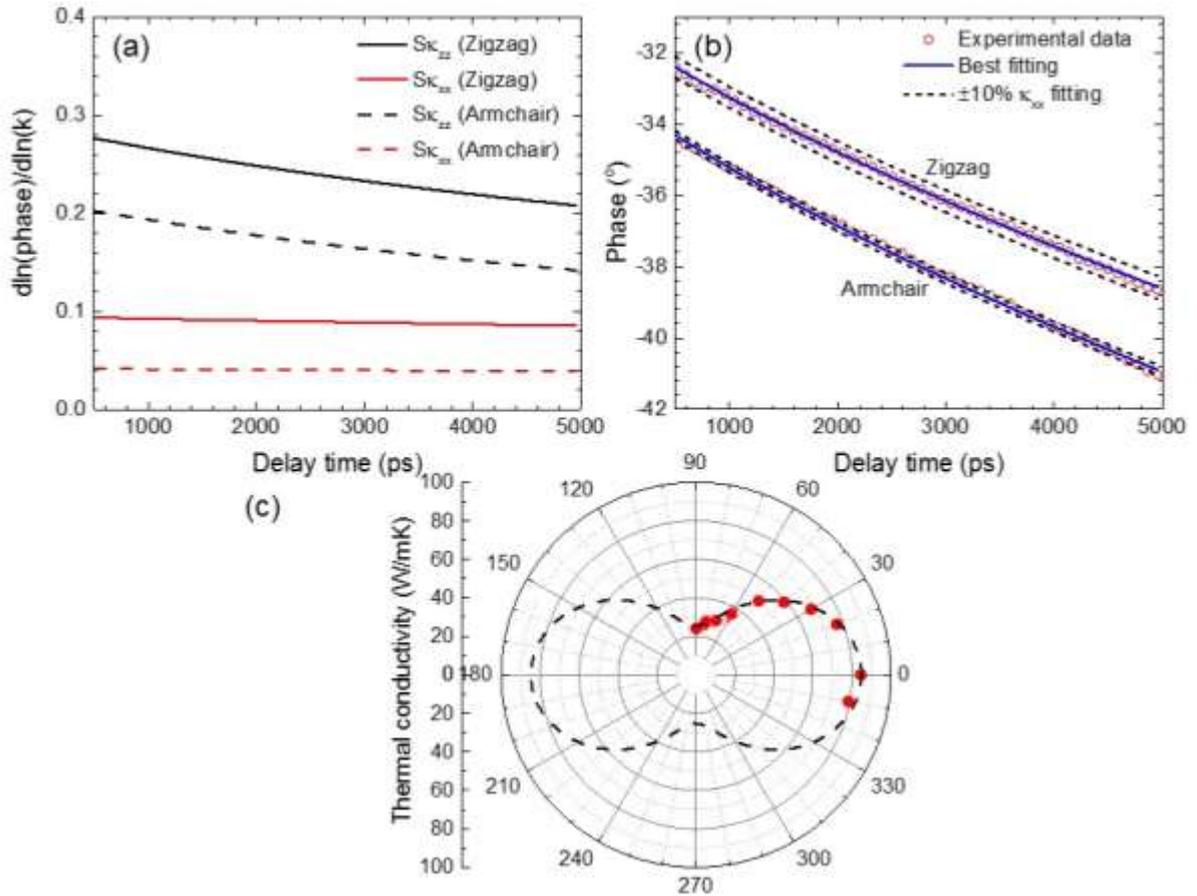

Figure 5. AB-TDTR measurement of anisotropic thermal conductivity of black phosphorus with angel dependence. (a) Sensitivity analysis for $\kappa_{xx}$ measurement when the major direction of elliptical beam aligned with armchair or zigzag directions. The modulation frequency is 1.1 MHz and spot size is $D_x = 10$ μm, $D_y = 300$ μm. (b) Two typical AB-TDTR phase data sets for in-plane thermal conductivity measurement along armchair and zigzag directions. (c) Angle dependent thermal conductivity of BP. The dash line is the theoretical prediction of thermal conductivity based on thermal conductivity of zigzag direction and armchair direction.

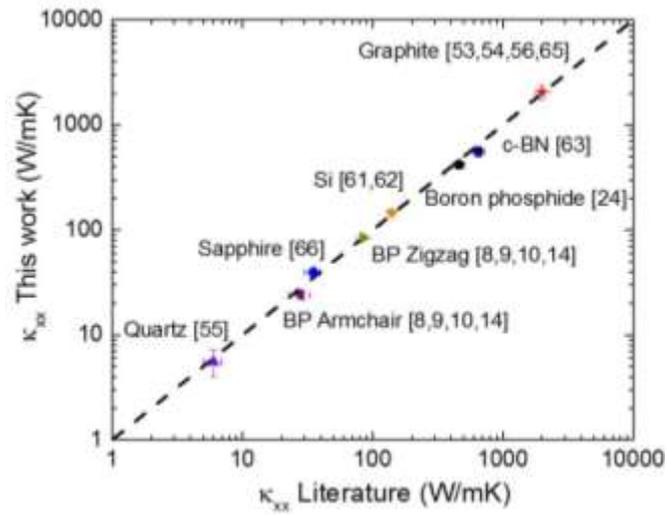

Figure 6. Summary of example measurement data from AB-TDTR versus literature reported values.

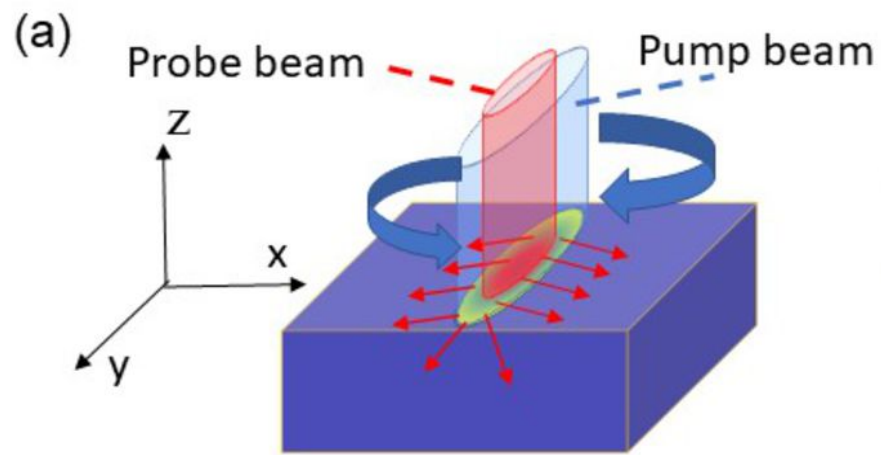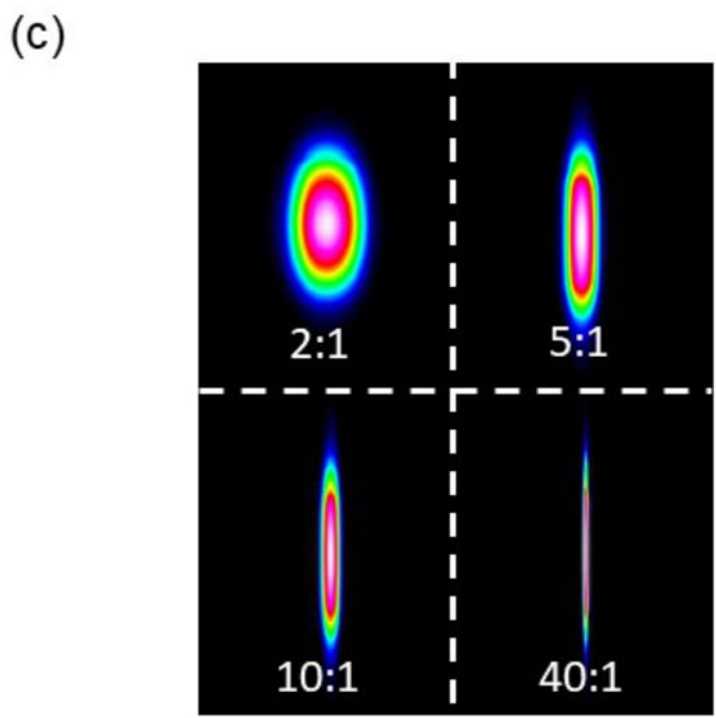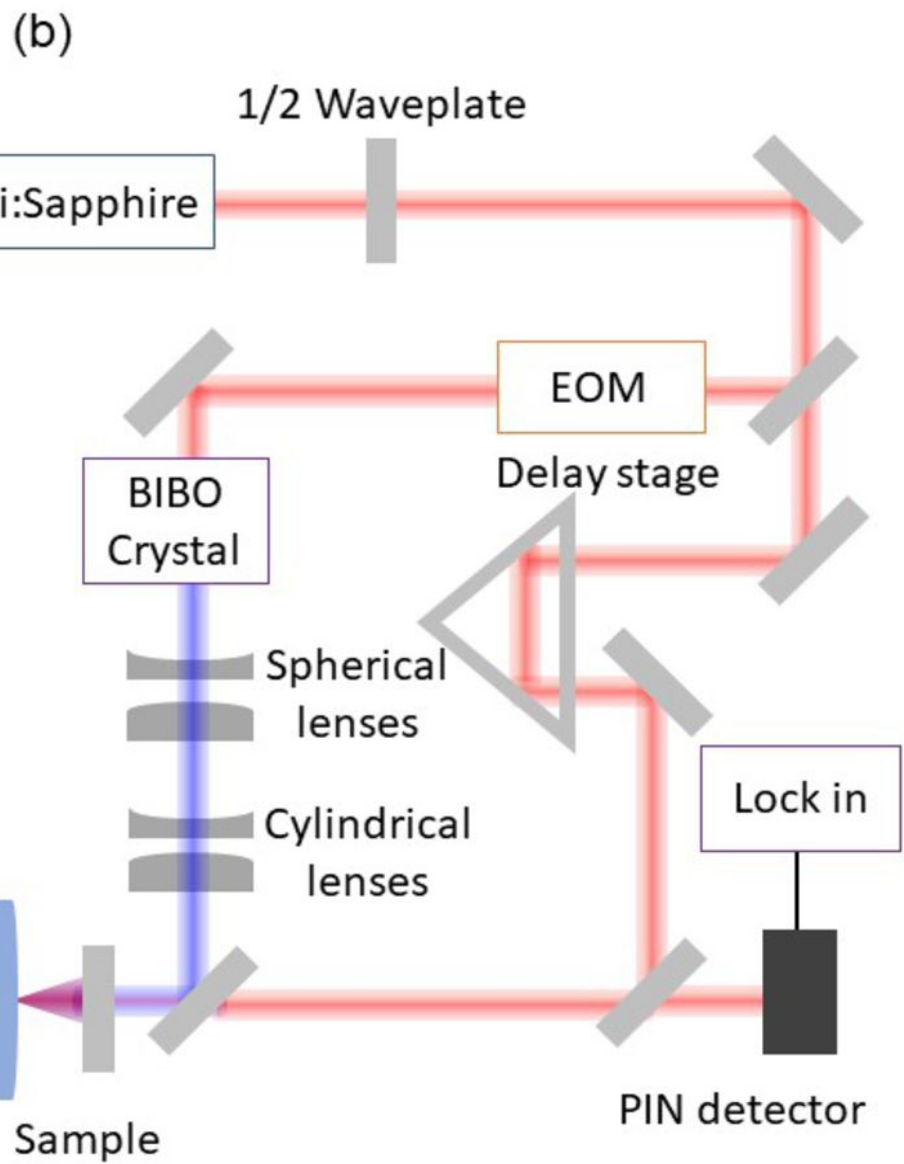

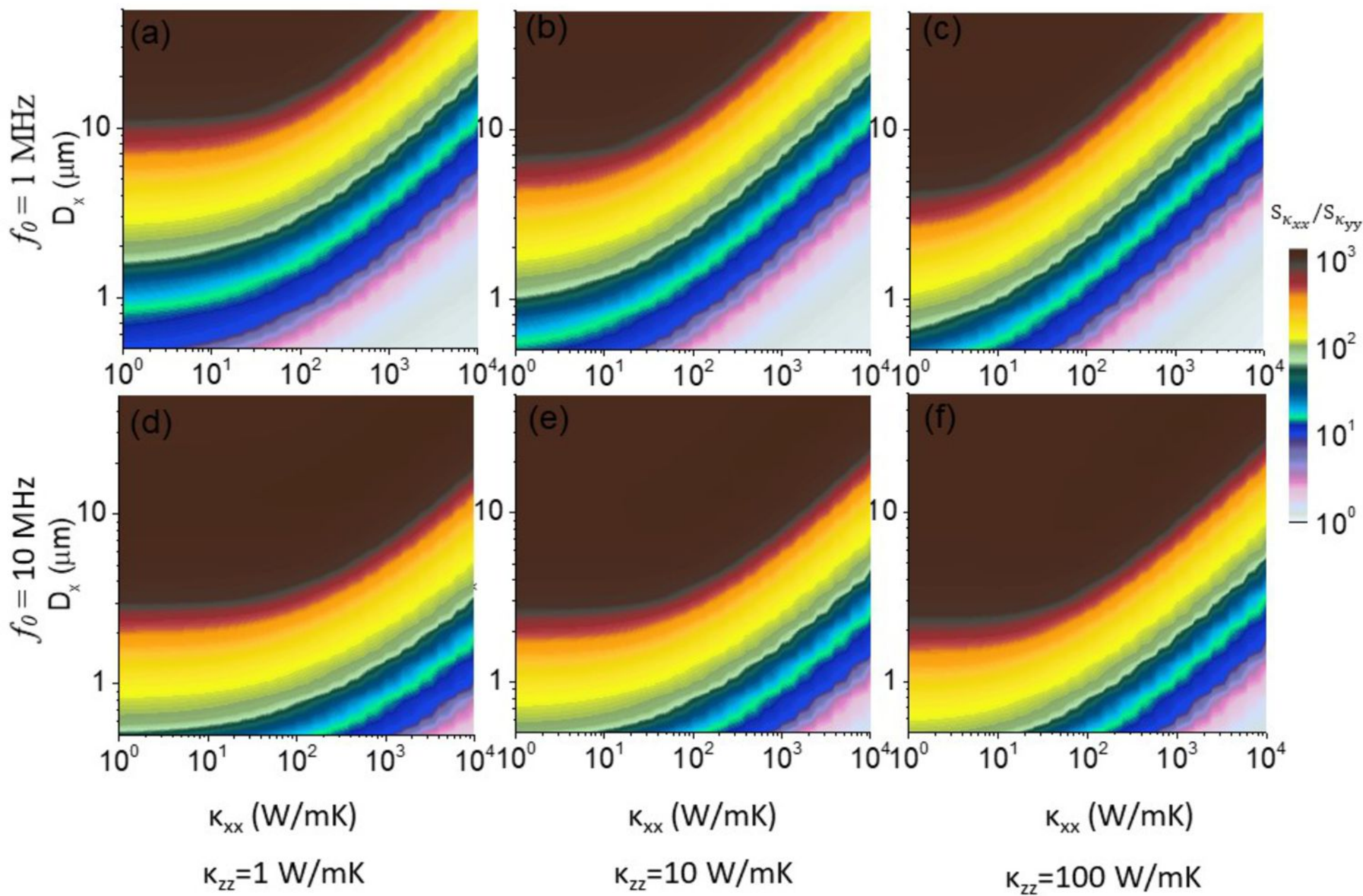

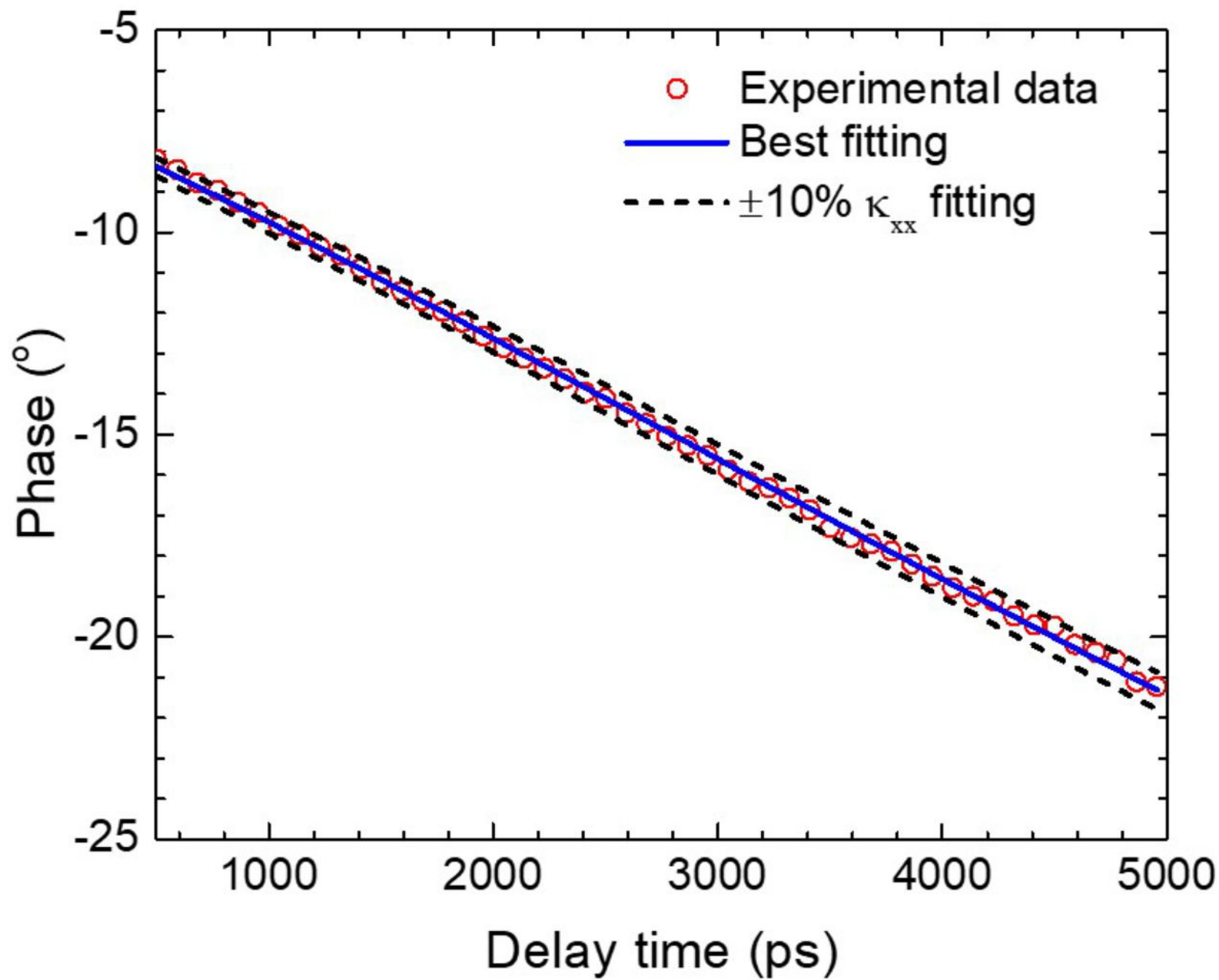

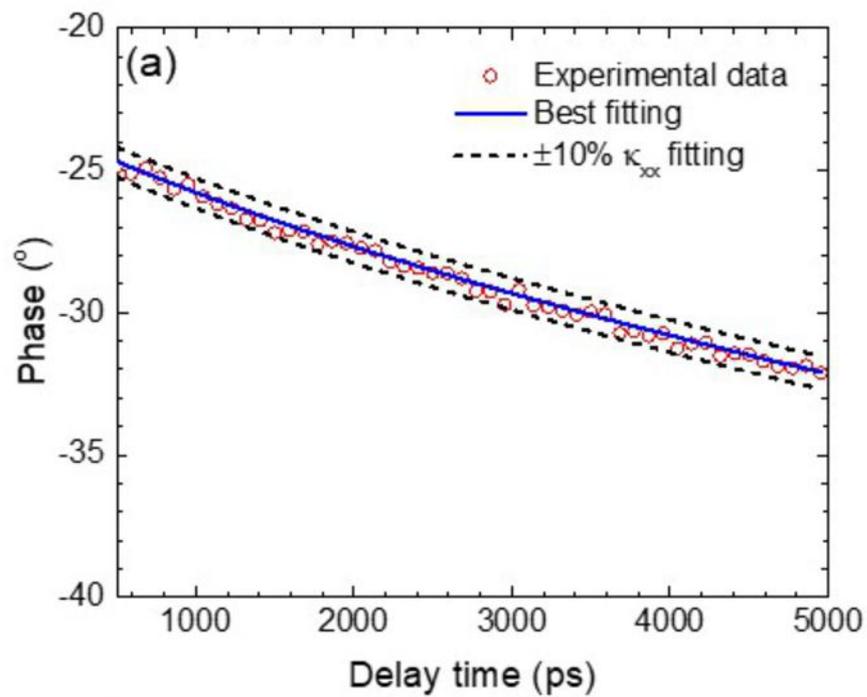
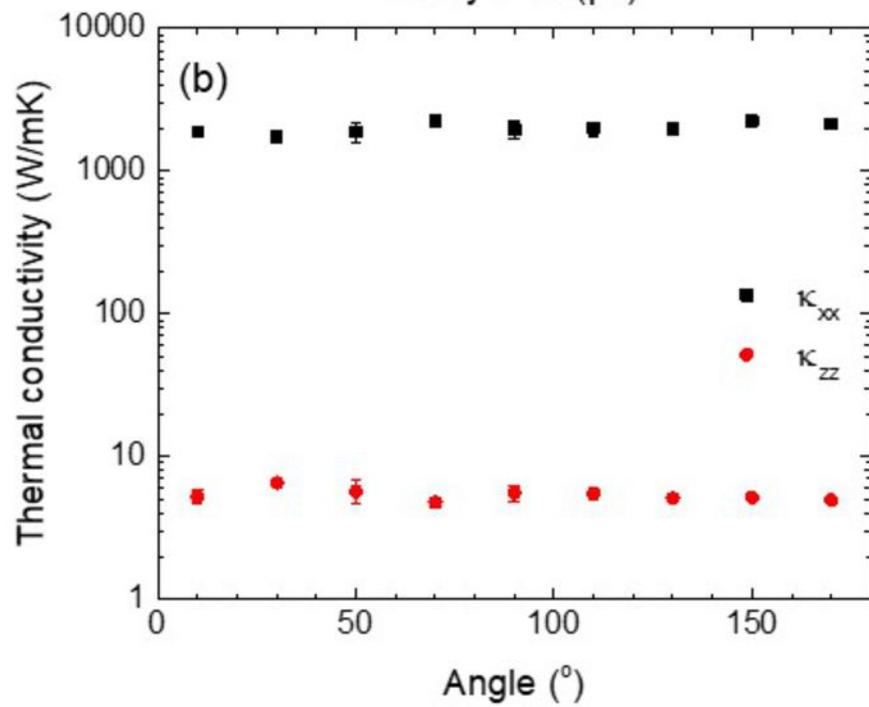

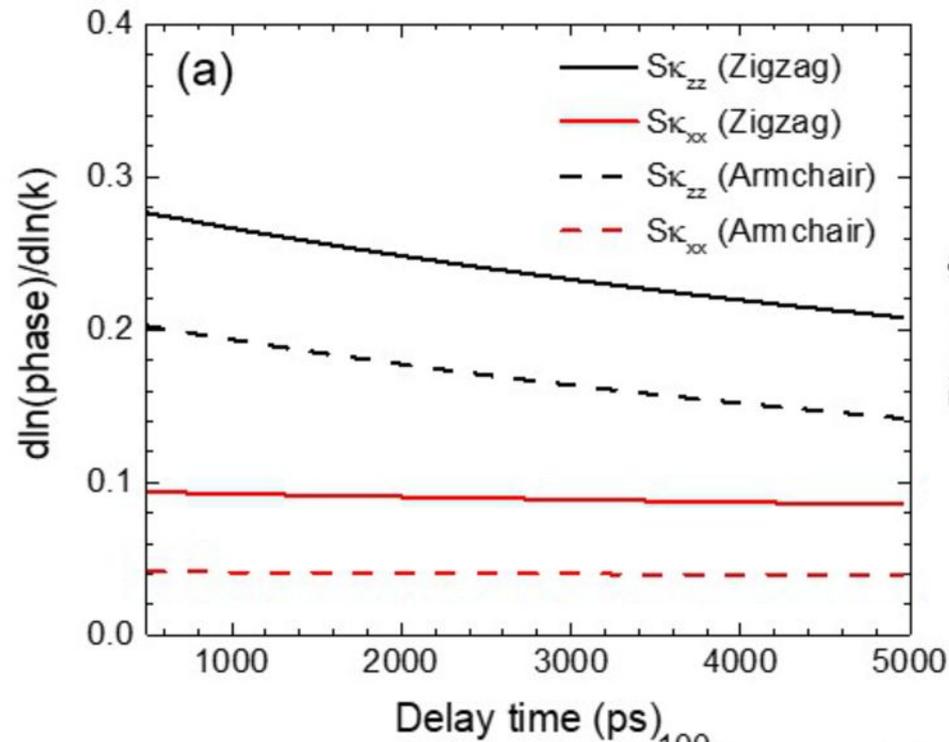
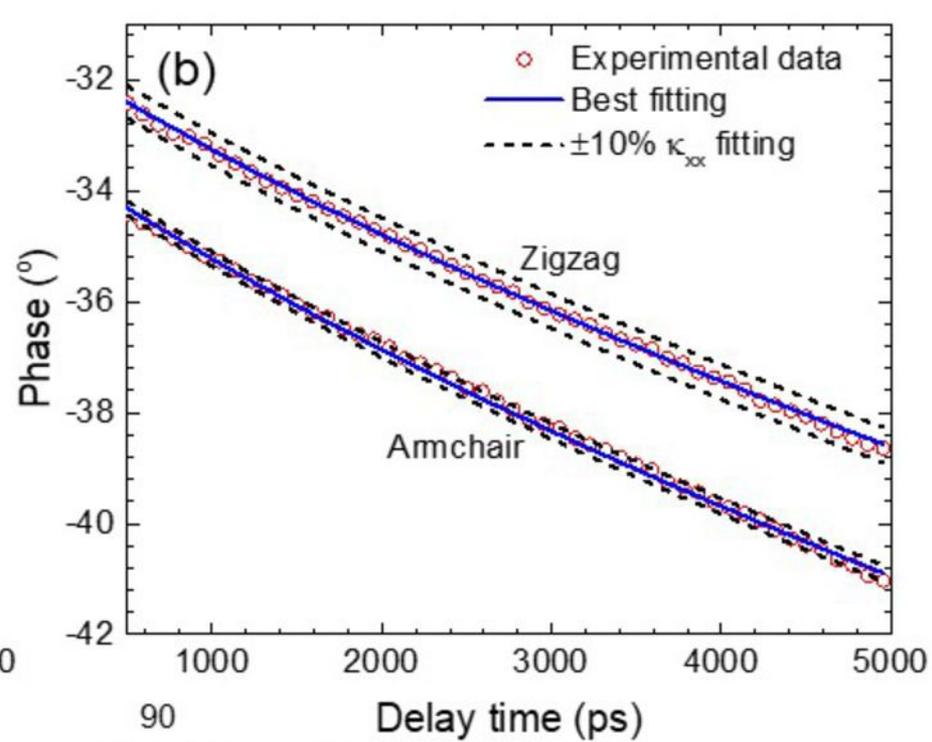
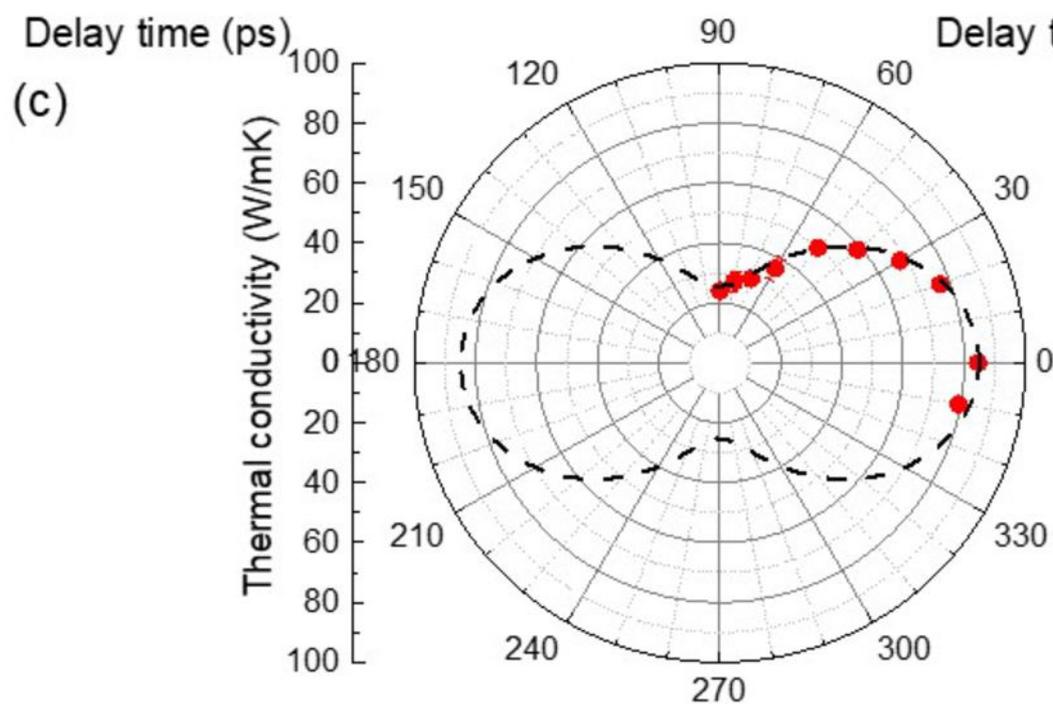

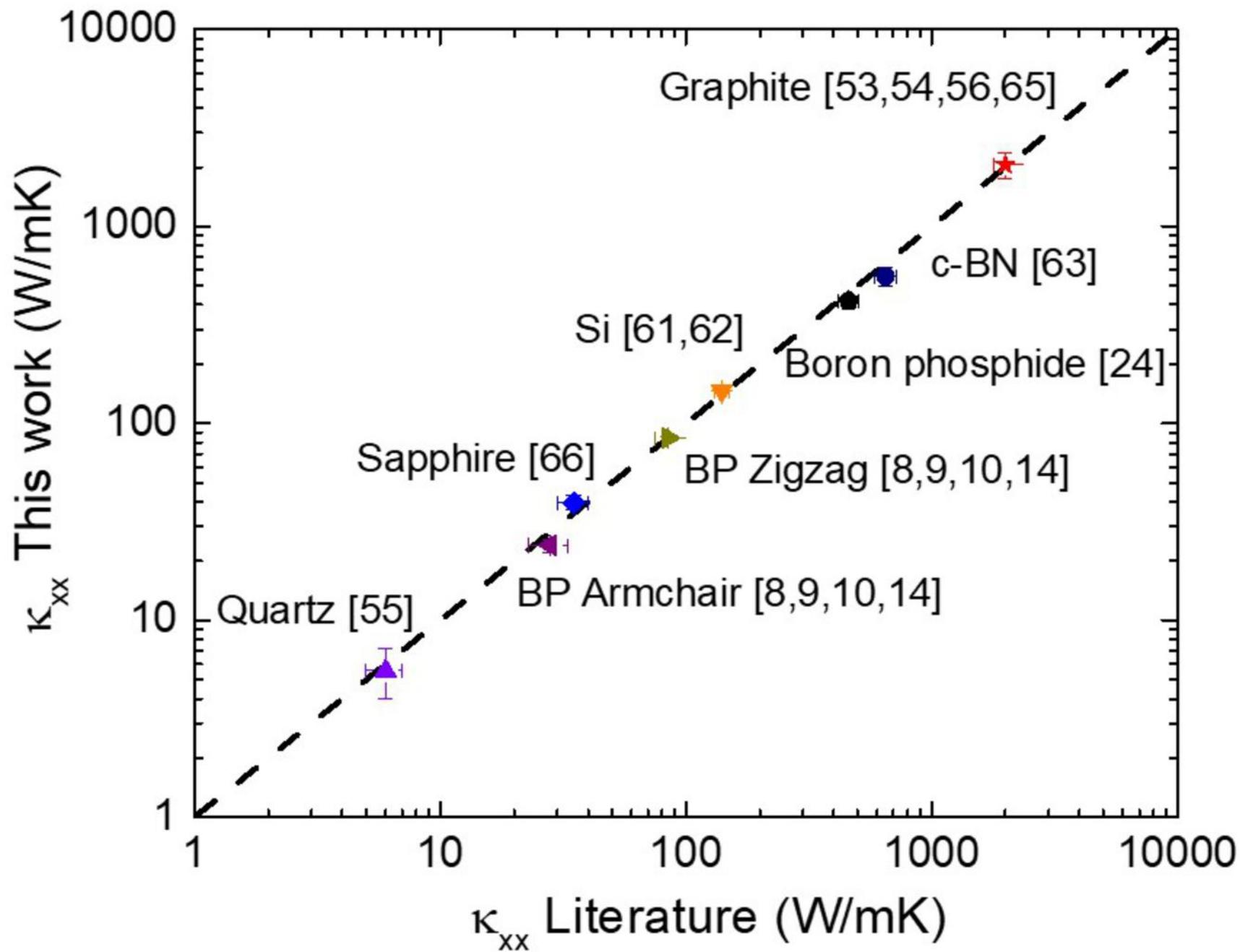